# Tuning flux-pinning in epitaxial $NdBa_2Cu_3O_{7-\delta}$ films via engineered, hybrid nanoscale defect structures

Sung Hun Wee[1,2] [*], Amit Goyal[1], Yuri L. Zuev[1] and Claudia Cantoni[1]

[1]*Materials Science and Technology Division, Oak Ridge National Laboratory, Oak Ridge, Tennessee 37831, USA*

[2]*Department of Materials Science and Engineering, University of Tennessee, Knoxville, Tennessee 37996, USA*

Epitaxial $NdBa_2Cu_3O_{7-\delta}$ films with a hybrid nanoscale defect structure comprised of $BaZrO_3$ nanodot arrays aligned along the *c*-axis in one half of the film thickness and aligned perpendicular to the *c*-axis in the other half thickness of the film were fabricated. Transmission electron microscopy images confirm the orientation of the nanoscale defect structures. The angular dependence of critical current density, $J_c$, at 77 K, 1 T, shows significantly reduced angular variation of $J_c$. This study nicely demonstrates how pinning characteristics can be tuned by tuning the nanoscale defect structures within the films.

---

[*] E-mail address: wees@ornl.gov

Defects or imperfections (chemical or structural) within high temperature superconductors such as $REBa_2Cu_3O_{7-\delta}$ (RE=rare earth elements including Y, REBCO) are energetically favorable sites to pin magnetic flux-lines (or vortices), the motion of which yields undesirable dissipation and limits the critical current density, $J_c$. Hence, understanding and controlling defect structures within REBCO coated conductors is key to realizing high $J_c$ with limited dissipation in the presence of high, applied magnetic fields. In the absence of extrinsically introduced defects, the $J_c$ of REBCO films is much higher for fields applied parallel to the *ab*-plane, *H//ab*, than for fields applied parallel to the *c*-axis, *H//c*. This anisotropy is due to the strong intrinsic pinning arising from the layered Cu-O structure. Significant research has therefore been directed at improving the pining for *H//c*. Attempts to intentionally introduce nanoscale defects in REBCO films range from growing films on substrates with surface-decorated nanoparticles[1,2] to incorporation of second phase nanoparticles in YBCO films.[3-5] In addition, it has been demonstrated that incorporation of a three dimensional (3D) self-assembled structure of $BaZrO_3$ (BZO), YSZ and $BaSnO_3$ nanodots and nanorods, resulting in a columnar structure similar to that formed by heavy-ion irradiation[6] without exposure to radiation, is a very effective way to enhance the pinning, particularly for *H//c*.[7-13] Such columnar defects resulted in strongly enhanced *c*-axis correlated pinning with pronounced peak in $J_c$ for *H//c*. In contrast, little enhancement in *ab*-correlated pinning is usually observed in these REBCO films with columnar defects.[11,12,14] As a result[12,14], some REBCO films even exhibited an "inverse anisotropy" with the highest $J_c$ at *H//c* and the lowest $J_c$ at *H//ab* in the temperature field regimes of 75-77 K, 1 T. In order to enhance $J_c$ for *H//ab*, controlled introduction of nanoscale defects which are aligned parallel to the *ab*-plane is needed. We have recently demonstrated control of self-assembly of BZO nanodots by controlling the strain in the growing film.[15] By modulating the strain along the vertical direction parallel to the *c*-axis of REBCO, it was shown that BZO nanodots can be aligned perpendicular to the *c*-axis of REBCO. Such films resulted in a strong peak in $J_c$ for *H//ab*. An idea motivated by these recent results is to fabricate hybrid REBCO films wherein half the film has defects structures optimized for *H//c* and the other half of the film has defect structures optimized for *H//ab*, in order to realize simultaneous improvement in flux-pinning for both applied field directions of *H//c* and *H//ab*. We report here on the successful fabrication of such a hybrid film and demonstrate that the anisotropy in $J_c$ is significantly reduced in this hybrid film. The REBCO film chosen for this work is $NdBa_2Cu_3O_{7-\delta}$ (NdBCO), but the strategy, constructing a hybrid defect structure should also expect to be effective to control angular anisotropy of $J_c$ for all other REBCO films. Hybrid NdBCO films consisted of two distinctive layers each comprising half the thickness of the film – (1) BZO-doped NdBCO layer (0.43 μm) and (2) multilayer of [$NdBCO_{3\ unit\ cell}$ /$BZO_{0.5\ unit\ cell}$] (0.42 μm). The samples were prepared on short segments (0.5 × 2 cm$^2$) of ion-beam-assisted deposition (IBAD) MgO templates with a $LaMnO_3$ (LMO) cap layer obtained from SuperPower Inc. The superconducting films were deposited using pulsed laser deposition with a KrF laser (λ=248 nm) at a repetition rate of 10 Hz. Laser energy density and substrate to target distance were 1.5 J/cm$^2$ and 7 cm, respectively. Films were grown at a substrate temperature ($T_s$) of 780$^o$C in 1%$O_2$/Ar gas at a deposition pressure of 800 mTorr. The bottom layer of NdBCO/BZO multilayer was grown by sequential ablation of pure NdBCO and BZO targets to form BZO nanodots aligned parallel to the *ab*-plane of the film. In this case the NdBCO spacer layer was thick enough so that the misfit strain from BZO nanodots did not get transmitted to the next layer.[15] This was followed by deposition of the second distinct half of the film comprising from a single NdBCO target containing 2 volume % BZO nanopowder, to produce 3D self-assembled stacks of BZO nanodot arrays aligned along the *c*-axis. After deposition, samples were *in-situ* annealed at $T_s$ = 500°C and $P(O_2)$ = 500 Torr. Ag electrodes were then sputtered onto the films followed by *ex-situ* annealing at 450°C for 1 h in flowing $O_2$ gas. Transport properties of the samples were measured using the standard four-probe method. Cross sectional microstructures were characterized by transmission electron microscopy (TEM).

Figure 1 shows the cross-section TEM images for the hybrid NdBCO film grown on the LMO/IBAD-MgO/Hastelloy template. Fig. 1(a) is a low magnification image showing the entire cross-section of the hybrid NdBCO film composed of two distinctive layers as divided by a white dot line. The top layer is NdBCO+BZO layer with the thickness of 0.43 μm and the bottom layer is NdBCO/BZO multilayer with the thickness of 0.42 μm. As shown in Fig. 1(b), a higher magnification image clearly shows a dark diffraction contrast representing BZO nanodots in the layers. In the top half of the film comprising a NdBCO+BZO layer, a dense and homogenous distribution of self-aligned BZO nanodots resulting in columnar defects aligned along the *c*-axis is observed.[7-14] The column widths are ~7-10 nm and the intercolumnar distance is ~20 nm, similar to what we have observed previously in BZO-doped NdBCO films.[11,12] In the NdBCO/BZO multilayer, BZO nanodot arrays are observed to have the long range alignment parallel to the *ab*-plane and should contribute to enhance the *ab*-correlated pinning.

The angular dependence of $J_c$ at 77 K, 1 T for the hybrid NdBCO sample is shown in figure 2. Here the magnetic field angle is recorded as the field is rotated away from the *c*-axis toward the *ab*-planes. The measurement was made in the maximum Lorentz force configuration in which the applied field direction is always perpendicular to the direction of current flow. $J_c$ data for pure NdBCO and NdBCO+BZO films with similar thickness reported ref. [12] are also included for comparison. The hybrid NdBCO film had a transport $J_c$ of 2.5 MA/cm$^2$ at 77 K, self-field which is higher than the $J_c$ of ~2 MA/cm$^2$ for NdBCO+BZO sample and the $J_c$ of 1.4 MA/cm$^2$ for pure NdBCO film. The NdBCO+BZO sample had strong "inverse anisotropy" in angular dependent $J_c$ with the highest $J_c$ at $H//c$. This was attributed to a massive improvement in $J_c$ for $H//c$ via *c*-axis aligned BZO nanodot columns and a relatively small increase in $J_c$ for $H//ab$ by collaterally enhanced random pinning resulted from misfit dislocations.[7,11,12] Compared to the NdBCO+BZO, the hybrid NdBCO film with self-assembly of BZO nanodots in both orientations – parallel to the *c*-axis and perpendicular to the *c*-axis or parallel to the *ab*-plane, significantly reduces the anisotropy in the angular dependence of $J_c$. As a result, a significantly higher minimum level of $J_c$, $J_c^{min}$, with respect to all field angles is obtained. The hybrid film was measured to have $J_c^{min}$ of 0.58 MA/cm$^2$ which is approximately 4 fold higher than the $J_c^{min}$ for a pure NdBCO sample (0.15 MA/cm$^2$) and also ~50% higher than that for NdBCO+BZO sample containing only *c*-axis oriented BZO nanodot columns.

Such an angular dependent behavior for the hybrid sample can be modeled using reference $I_c$ data obtained from NdBCO samples containing only singly oriented BZO nanodot arrays, either aligned parallel to the *c*-axis or parallel to the *ab*-plane, as shown in figure 3.  For this calculation, a NdBCO/BZO multilayer sample with ~0.6 μm thickness was also grown using identical procedures as for NdBCO/BZO multilayer in the hybrid sample.  A detailed experimental and theoretical study on series of NBCO/BZO multilayer samples with various thicknesses for the NdBCO layer and the BZO pseudo-layer will be reported elsewhere.[15]  The curve with a solid orange line represents the simulated angular dependent $I_c$ for the hybrid sample calculated from reference $I_c$ data of each of the NdBCO+BZO and NdBCO/BZO multilayer samples which are also shown in figure.  This curve is quite consistent with the experimentally measured $I_c$ for the hybrid sample shown in red.  The slight discrepancy arises because the 0.42 μm thick NdBCO/BZO bottom layer in the hybrid film in all likelihood has a higher $J_c$ than that of the 0.6 μm thick reference film used for modeling or calculations.  The same is also true for the upper portion of the hybrid film.  Exponential decay of $J_c$ with film thickness is well established.[16]  Despite the small discrepancy between modeled and experimentally measured data for the hybrid sample, these results demonstrate the feasibility of tuning the desired pinning characteristics by tuning the defect structures within the REBCO films.

In summary, a hybrid NdBCO film with one half of the film having insulating nanoscale defect structures aligned along the *c*-axis and the other half of the film having nanoscale defect structures aligned along the *ab*-plane was shown possible.  This unique hybrid nanoscale structure resulted in less anisotropy in the angular dependence of $J_c$ with significantly improved $J_c^{min}$, compared to pure NdBCO films and those with nanoscale defects aligned along only one direction.

**ACKNOWLEDGMENTS**

S. H. Wee and Y. L. Zuev would like to thank Oak Ridge Associated Universities for a postdoctoral fellowship. The authors would like to thank Sy Cook at Oak Ridge National Laboratory for laser-scribing a microbridge on the sample. The authors also thank V. Selvamanikcam at SuperPower Inc. for providing the Hastelloy substrates with an IBAD MgO layer / Homoepitaxial MgO layer / Epitaxial $LaMnO_3$. Research was sponsored by the US Department of Energy, Office of Electricity Delivery and Energy Reliability- Superconductivity Program, under contract DE-AC05-00OR22725 with UT-Battelle, LLC managing contractor for Oak Ridge National Laboratory.

**FIGURE CAPTIONS**

Figure 1. Cross-section transmission electron micrographs of 0.85 μm thick, hybrid NdBCO film grown on IBAD-MgO templates with a hybrid nanoscale defect structure. (a) Full cross-sectional area of entire NdBCO layer on IBAD-MgO templates with configuration of $LaMnO_3$ / MgO / $Al_2O_3$ / Hastelloy. (b) High magnification image of hyrid NdBCO film showing a high density of columnar defects of $c$-axis oriented BZO nanodots in the top layer of NdBCO+BZO, and a long range order of BZO nanodot arrays aligned the $ab$-plane in the bottom layer of NdBCO/BZO.

Figure 2. Angular dependence of $J_c$ at 77 K and 1 T for 0.85 μm thick, hybrid NdBCO film, in a magnetic field applied -30° to 110° with respect to the $c$-axis of the film. For comparison, the $J_c$ data for pure NdBCO and NdBCO+2 vol.% BZO samples reported from ref. [12] are also illustrated.

Figure 3. Angular dependence of $I_c$ at 77 K and 1 T for 0.85 μm thick, hybrid NdBCO film measured and modeled using reference $I_c$ data obtained from each of NdBCO samples only containing BZO nanodot arrays aligned to either the $c$-axis or the $ab$-plane, which are also shown in the figure.

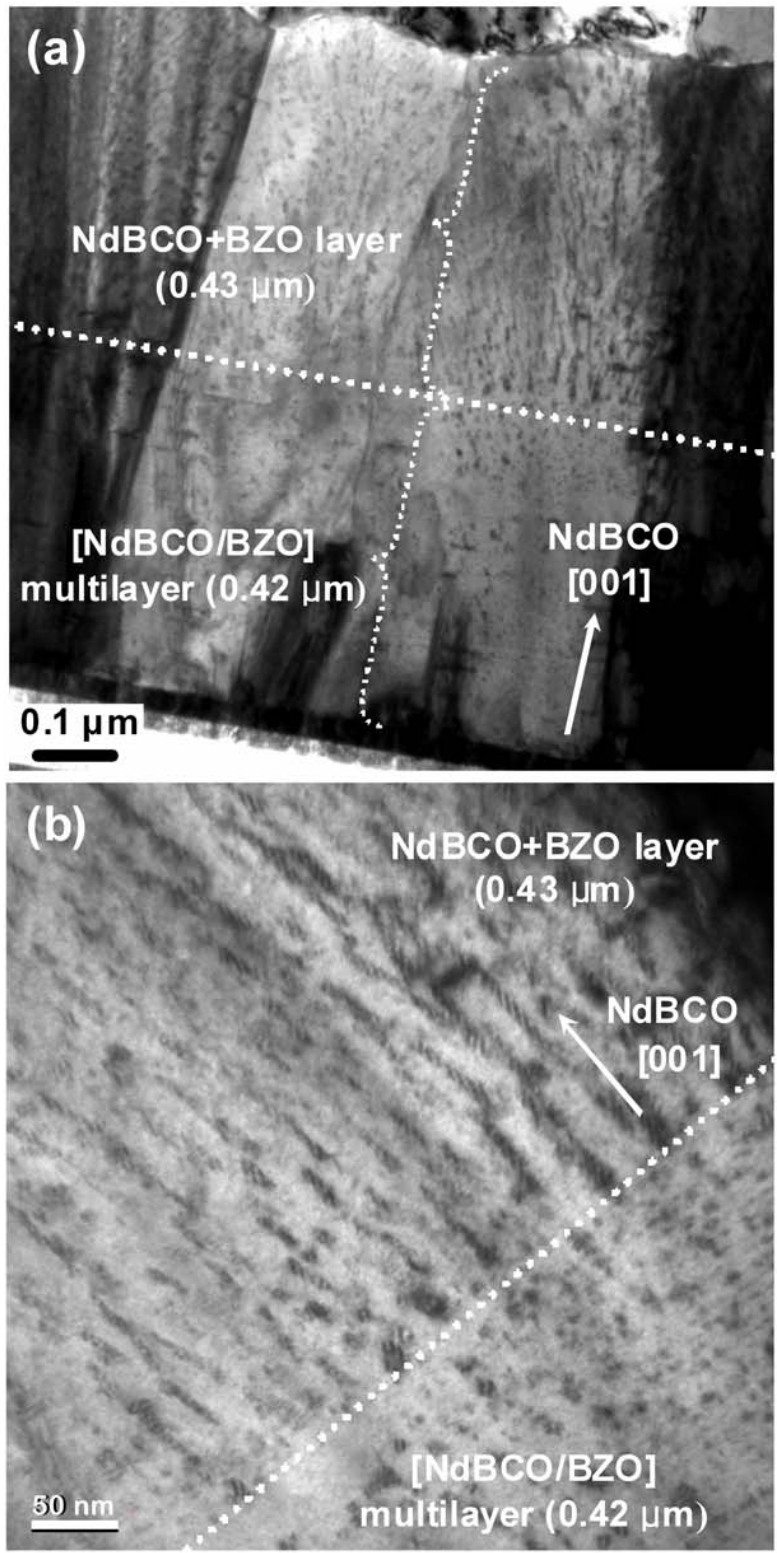

Figure 1

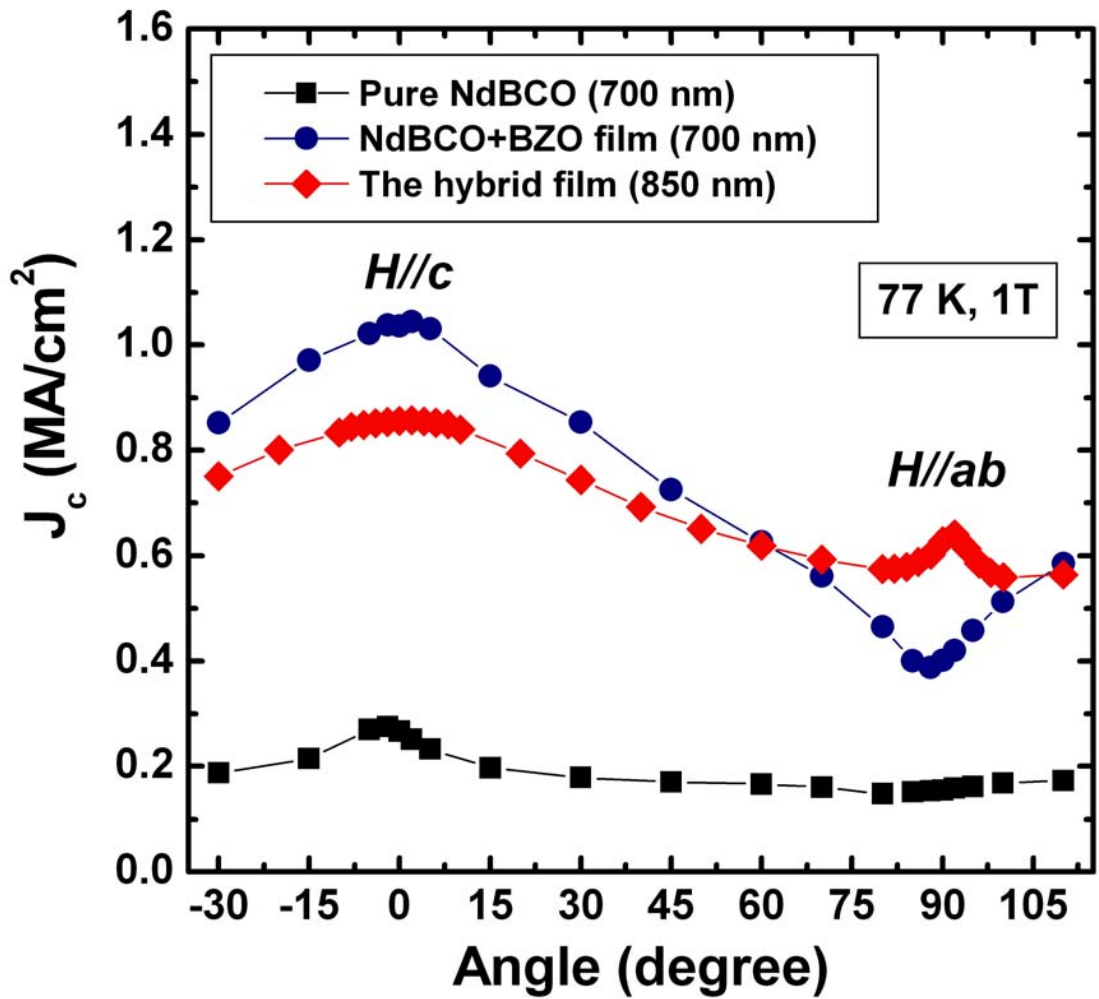

Figure 2

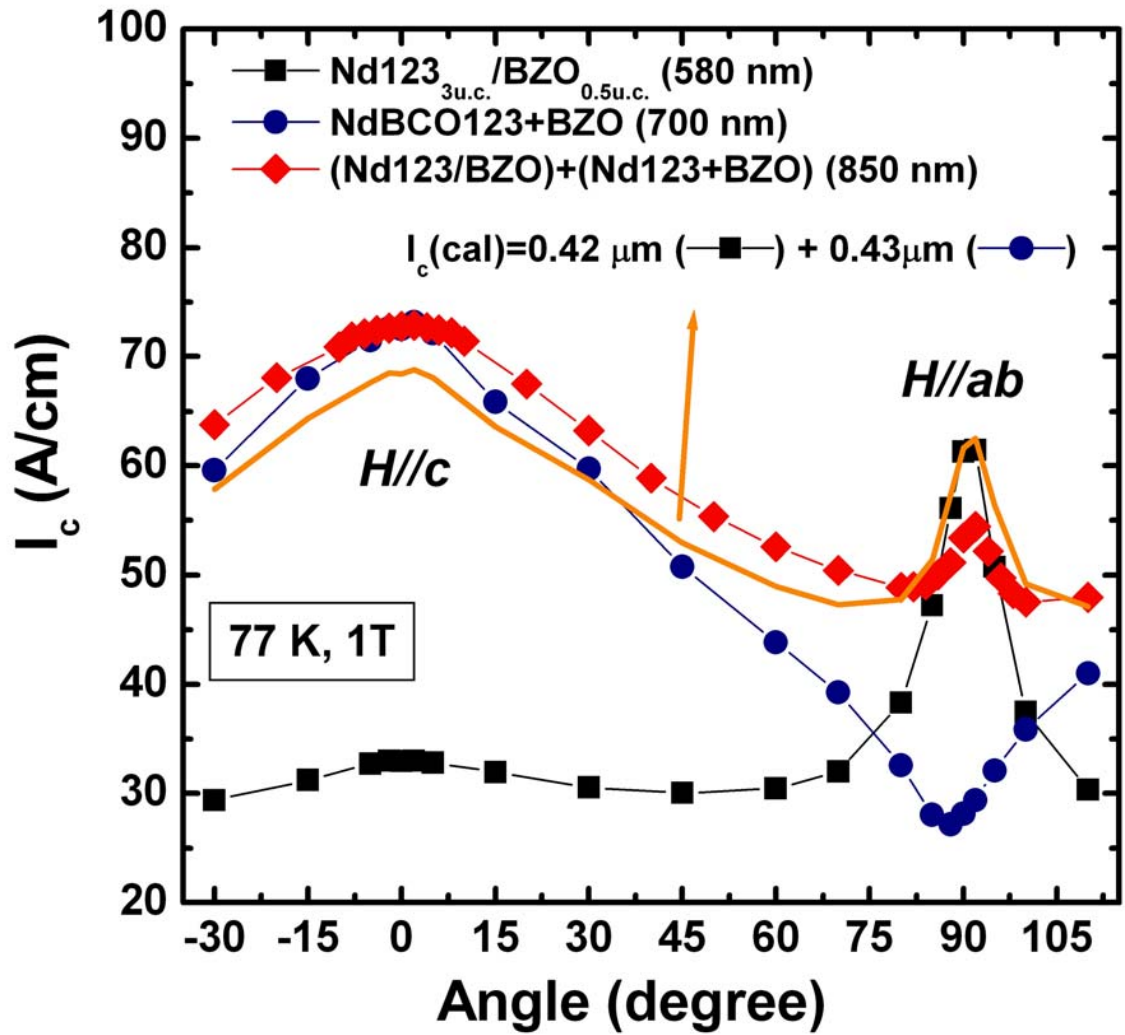

Figure 3